\input ppltexa.sty
\input epsf
% Revised version (Sept. 1980)
% Original version finished on May 26, 1980
% To be processed with TEX 112 and TEMPOR REFSA

\let\rjustline\rightline
\let\eqv\equiv
\def\twocol{\matrix2}
\def\eqaligntwo{\eqalignn2}
\let\xdot\dot \let\dot\cdot
\def\^{\hat}\def\.{\xdot}
\def\Escr{{\cal E}}
\def\Fscr{{\cal F}}
\def\Mscr{{\cal M}}
\def\Pscr{{\cal P}}
\def\ldotss{\ldots{}}
\let\ital\it
\hsize 5.5 in \hoffset=0.5in
\vsize 8 in \def\pagesize{8.5 in} \voffset=0.0in
\long\def\endpaper{\ifempty\refsfile\else\par\input\refsfile\fi\par\vfill\end}
\def\ref #1{$^{\hbox{\sevenrm #1}}$}

\defrefsfile{temporref.sty}
\sections
\qdef\intro \qdef\deriv \qdef\basic \qdef\numerics
\qdef\thresh  \qdef\conclusions
\appendices \qdef\appnum
\endpreamble

\rjustline{PPPL--1672 (June 1980)}
\rjustline{Phys.\ Fluids \vol{24}(1), 127--137 (Jan.~1981)}
\vskip 0.75 in
\hbox{\titl Temporal evolution of lower hybrid waves in the presence}
\hbox{\titl of ponderomotive density fluctuations}
\vskip 20 pt
\hbox{\hskip 20pt Charles F. F. Karney}
\vskip 5 pt
\hbox{\hskip 20pt Plasma Physics Laboratory, Princeton University,}
\hbox{\hskip 20pt Princeton, New Jersey 08544}
\vskip 5 pt
%\hbox{\hskip 20pt (Received }

\abstract

The propagation of lower hybrid waves in the presence of
ponderomotive density fluc\-tu\-a\-tions is considered.  The problem is
treated in two dimensions and, in order to be able to correctly
impose the boundary conditions, the waves are allowed to evolve in
time.  The fields are described by
$iv_\tau - \int v_\xi\,d\zeta + v_{\zeta\zeta} + \abs v ^2 v = 0$
where $v$ is proportional to the electric field, $\tau$ to time, and
$\zeta$ and $\xi$ measure distances across and along the lower
hybrid ray.  The behavior of the waves is investigated numerically.
If the amplitude of the waves is large enough, the spectrum of the
waves broadens and their parallel wavelength becomes shorter.  The
assumptions made in the formulation preclude the application of
these results to the lower hybrid heating experiment on
Alcator--A.  Nevertheless, there are indications that the physics
embodied in this problem are responsible for some of the results
of that experiment.

\section \intro.  Introduction.

The injection of rf power near the lower hybrid
frequency is an attractive method for the auxiliary
heating of tokamak plasmas.\ref\stix\  Because of the
high powers required (several MW) and because
lower hybrid waves principally propagate along
well-defined resonance cones,\ref{\kuehla--\briggs}
there has been considerable interest in nonlinear effects
on the propagation of lower hybrid waves.  This
problem was first addressed by Morales and Lee\ref\moralesa\
who studied the two-dimensional elec\-tro\-sta\-tic propagation
of one of the two lower hybrid rays in a homogeneous
plasma.  Although this is perhaps the simplest model
that can be considered, a correct treatment of this
problem has yet to be made.  It is this deficiency
that this paper attempts to remedy.

Briefly the difficulty of this problem arises as follows:
If it is assumed that the rf fields in the plasma have
reached a steady state, i.e., that the potential is
given by $\phi(x,z)\exp(-i\omega_0 t)$ ($x$ and $z$ are
coordinates perpendicular and parallel to the ambient magnetic
field $B_0$), then the electric field
obeys the complex modified Korteweg--deVries
equation.\ref{\moralesa--\karney}  This equation
is mathematically well-posed when solved as an initial
value problem in one of the coordinates $x$, that is when
$\phi(x=0,z)$ is given.  Unfortunately, this does not
correspond to physically realizable boundary
conditions since waves can propagate in both the $+x$ and
$-x$ directions in a single ray.\ref{\karney, \bellan}  When the
correct boundary conditions are imposed, there is
numerical evidence that solutions of the complex modified
Korteweg--deVries equation need not exist and that this equation
is therefore ill-posed.\ref\karney\  This is confirmed by our
finding solutions inconsistent with the assumption
of a steady state (see Sec.\ \numerics).
The problem arises because the direction of power flow
which determines how to impose the boundary conditions is
defined only with reference to a problem in which a temporal
evolution of the wave packet is allowed.  In assuming a 
steady state for the electric field amplitude, the equation
no longer has built into it the crucial ingredient which
determines how the boundary conditions are imposed.  This
defect is corrected by including a slow time dependence of the
potential [so that the potential is given by
$\phi(x,z,t)\exp(-i\omega_0 t)$].
This then leads to a nonlinear partial differential
equation in two spatial dimensions and time, which we will
study numerically in this paper.

In formulating this problem, we shall ignore many effects which
should possibly be included to obtain a complete understanding of
the propagation of lower hybrid waves.  This will enable us to study
the effect of the nonlinearity in as simple a system as possible.
Even so, the numerical solution of a partial differential equation
in two dimensions and time is time consuming and, as we shall see,
the behavior of the fields as described by this equation can be
quite complicated.  In the absence of any analytical methods for
solving this equation, we must therefore be content with the
solution for only a few cases.  This will enable us to confirm the
threshold for strong nonlinear effects given in Ref.\ \karney\ and
to give a more complete description of the nature of these
nonlinear effects.  Because of the approximations made
in formulating the problem, we shall not be able to
apply these results to the propagation of lower hybrid waves
near the edge of a tokamak plasma where nonlinear effects are most
strong.  However, there are indications that the physical processes
that are considered in this paper do play a role in the propagation
in that region.

The plan of this paper is as follows:  In Sec.\ \deriv\ we
derive the partial differential equation governing the
temporal evolution of $\phi$.  The basic properties of
this equation will be discussed in Sec.\ \basic.  In particular,
we will show how the right boundary conditions automatically
drop out from the equation.  Since the equation is analytically
intractable, we resort to a numerical integration, the
results of which are given in Sec.\ \numerics.  The threshold
condition for a strong nonlinear interaction is derived in
Sec.\ \thresh.  The results
are summarized and the consequences to lower hybrid heating
of tokamaks are presented in Sec.\ \conclusions.

\section \deriv.  Formulation of the Problem.

In this section, we derive the partial differential equation
governing the evolution of the electric field for one
of the lower hybrid rays in two spatial dimensions and time.
We assume that the fields are elec\-tro\-sta\-tic, that the
plasma is homogeneous, and that it is immersed in a uniform magnetic
field.  The derivation closely follows that of the complex modified
Korteweg--deVries equation by Morales and Lee;\ref\moralesa\
the additional ingredient we consider is the slow time dependence
of $\phi$.

The potential of the lower hybrid wave is taken to have
the form
$$\Re[\phi(x,z,t)\exp(-i\omega_0 t)]$$
where the $t$ dependence of $\phi$ is taken to be much
slower than $\omega_0$.  Since we are only
doing the two-dimensional problem, we have $\partial/\partial y
\eqv 0$.  In the elec\-tro\-sta\-tic limit,
$\phi$ obeys Poisson's law
$$\del\dot\mat K(\del, \partial/\partial t, \abs{\del \phi}^2)
\dot\del\phi = 0. \eqn(\en\poisson)$$
Here we regard the dielectric tensor $\mat K$ as an operator
through its arguments $\del$ and $\partial/\partial t$.  The
dependence on $\abs{\del \phi}^2$ accounts for the nonlinearity.
If the temporal evolution of $\phi$ is sufficiently slow (compared
with ion acoustic time scales), then
nonlinearities due to ponderomotive density changes can
be written in this way.  Parametric instabilities are excluded
from our consideration.

If the dependence of $\phi$ on $x$, $z$, and $t$ is weak and
if the $\phi$ itself is small so that the nonlinearity is
weak, we may expand $\mat K$ to obtain
$$\eqalignno{
\mat K(\del, \partial/\partial t, \abs{\del \phi}^2)=\mat K
&+{1\over2}\epsilon{\partial^2\mat K\over\partial\del\partial\del}
\dotdot\del\del\cr
&+\epsilon{\partial\mat K
\over\partial(\partial/\partial t)}{\partial\over\partial t}
+\epsilon{\partial\mat K\over\partial\abs{\del \phi}^2}
\abs{\del \phi}^2+O(\epsilon^2).
&(\en\kexpand)\cr}$$
All the evaluations of $\mat K$ on the right hand side of Eq.\ (\+)
are at the point $\del=\partial/\partial t= \abs{\del \phi}^2=0$.
The term involving $\partial \mat K/\partial\del$ is zero
because of the symmetries of a stationary plasma.
We have introduced a formal expansion parameter $\epsilon$ to
aid in the ordering of terms.  The fact that the last three
terms in Eq.\ (\+) are taken to be of order $\epsilon$ results
a maximal ordering.  If, in fact, one of the terms is much
smaller than the others, a subsidiary ordering can
be introduced to eliminate that term.

Since it is more usual to write $\mat K$ as a normal function,
rather than as an operator, we
rewrite $\mat K$ as $\mat K(\vec k, \omega, n)$ where
$\vec k$ is the Fourier-transform variable conjugate
to space, $\omega$ is that conjugate to time, and $n$
is the plasma density.  [A space-time dependence of
$\exp(i\vec k\dot\vec r - i\omega t)$, where $\vec r
= (x,z)$, is assumed.]  Then the derivatives in Eq.\ (\+)
become
$${\partial^2\mat K\over\partial\del\partial\del}=
-{\partial^2\mat K\over\partial\vec k\partial\vec k},
\qquad
{\partial\mat K\over\partial(\partial/\partial t)}=
i{\partial\mat K\over\partial\omega},
\qquad
{\partial\mat K\over\partial\abs{\del \phi}^2}=
{\partial n\over\partial\abs{\del \phi}^2}
{\partial\mat K\over\partial n}.$$
In this case, all the evaluations on the right hand sides
are performed at $\vec k=0$, $\omega = \omega_0$, and
$n = n_0$, where $n_0$ is the unperturbed density of the
plasma (i.e., in the absence of any electric fields).

We now write down Eq.\ (\poisson) to various orders in $\epsilon$.
To order $\epsilon^0$, we find
$$\del\dot\mat K(0, \omega_0, n_0)\dot\del\phi = 0 \eqn(\en\pzero)$$
where
$$\mat K(0, \omega_0, n_0) = \twocol{K_\perp&0\cr 0&K_\para\cr}.$$
Assuming that $K_\perp K_\para$ is negative,
Eq.\ (\pzero) is hyperbolic and its solution is
$$\phi = \phi_+(x-gz)+\phi_-(x+gz)$$
where $g = (-K_\perp/K_\para)^{1/2}$.

In general, both the right- and left-going rays will be present.
However, if the source of the lower hybrid rays is localized,
the two rays will be separated far from the source and in that
region we may then treat each ray in isolation.  Similarly, only
a single ray will be present if the source is adjusted so that
only that ray is excited.  We may therefore
assume that, to zeroth order, only the right-going ray is
present, i.e., $\phi_-=0$.  To order $\epsilon^1$, we
write $\phi=\phi(\epsilon t^\prime,\epsilon x^\prime, z^\prime)$
where $t^\prime = t$, $x^\prime=x$, $z^\prime=cz-sx$, and
$s=(1+g^2)^{-1/2}$, $c=gs$.  We should think of $x^\prime$ and
$z^\prime$ as measuring distances along and across the lower
hybrid ray.  Substituting this form of $\phi$
into Eq.\ (\poisson), we obtain to order $\epsilon^1$
$$iA{\partial^2E\over\partial z^\prime\partial t^\prime}
-B{\partial E\over\partial x^\prime}
+C{\partial^3E\over\partial z^{\prime3}}
+D{\partial\over\partial z^\prime}(\abs E^2 E)=0\eqn(\en\pone)$$
where $E=\partial\phi/\partial z^\prime$ is the electric field
measured in the direction normal to the lower hybrid ray,
$A$, $B$, $C$, and $D$ are real coefficients given by
$$\eqaligntwo{
A&=\^{\vec k}\dot{\partial\mat K\over\partial\omega}\dot\^{\vec k},&
B&=2sK_\perp,\cr
C&=-{1\over2}\^{\vec k}\^{\vec k}\dotdot
{\partial^2\mat K\over\partial\vec k\partial\vec k}
\dotdot\^{\vec k}\^{\vec k},&
D&=-{1\over n}{\partial n\over\partial\abs{\del\phi}^2},\cr}$$
and $\^{\vec k}$ is the unit vector $(-s,c)$.

These expressions may be evaluated using
the expression for $\mat K$ for a stationary Maxwell\-ian
plasma.\ref\akhiezer\
Usually the frequency of the incident rf power satisfies
$\Omega_i^2 \lsls \omega_0^2 \lsls \Omega_e^2$
where $\Omega_j$ is the cyclotron frequency of species $j$
($j=i$ or $e$ for ions or electrons).  In that case, the components
of the zeroth-order dielectric tensor may be written
$$K_\perp = 1+{\omega_{pe}^2\over\Omega_e^2}
-{\omega_{pi}^2\over\omega_0^2},
\qquad K_\para = 1-{\omega_{pe}^2\over\omega_0^2}
-{\omega_{pi}^2\over\omega_0^2},$$
where $\omega_{pj}$ is the plasma frequency for species $j$ with
density $n_0$.  The coefficients in Eq.\ (\pone) become
$$\eqalign{
A&={2\over\omega_0}
\left(1+{\omega_{pe}^2+\omega_{pi}^2\over\Omega_e^2}\right),\cr
C&=3{\omega_{pi}^2v_{ti}^2\over\omega_0^4}
+{3\over4}s^4{\omega_{pe}^2v_{te}^2\over\Omega_e^4}
-s^2c^2{\omega_{pe}^2v_{te}^2\over\omega_0^2\Omega_e^2}
+3c^4{\omega_{pe}^2v_{te}^2\over\omega_0^4},\cr
D&={1\over4}{\epsilon_0\over n_0(T_e+T_i)},\cr}$$
where $v_{tj}^2=T_j/m_j$, $T_j$ is the temperature of species $j$,
and $\epsilon_0$ is the dielectric
constant of free space.  In writing the expression for $D$,
we have used the formula for the density depression due
to the ponderomotive force.\ref{\moralesa, \litvak, \hsuan}
Note that $C$ is always positive.  The fact that $C$ and $D$ have the
same signs has an important effect on the propagation of the
lower hybrid waves.

We now rescale the variables in Eq.\ (\pone):
$$\xi=x^\prime/(B\sqrt C),\quad
\tau=t^\prime/A,\quad
\zeta=z^\prime/\sqrt C,\quad
v=\sqrt D\, E.\eqn(\en\normb)$$
The equation for $v$ is then
$$iv_\tau - \int_{-\infty}^\zeta v_\xi\,d\zeta
+v_{\zeta\zeta} + \abs v ^2 v = 0\eqn(\en\eq a)$$
where subscripted Greek letters denote differentiation.  When
integrating Eq.\ (\pone) to give Eq.\ (\eq a), we assume that
$v$ and all its derivatives are zero at $\zeta=-\infty$ (i.e.,
far from the ray).
The accessibility condi\-tion\ref\golant\ imposes an auxiliary
condition on $v$ namely
$$\int_{-\infty}^\infty v \,d\zeta =0;\eqn(\+b)$$
in other words, the electric potential is equal at $\zeta=
\pm\infty$.
Equation (\+) describes the evolution of the electric
field of a single lower hybrid ray in two dimensions and
time under the influence of thermal dispersion and
nonlinear ponderomotive effects.
Equation (\eq a)
is the same as Eq.\ (50) of Ref.\ \karney\ if the replacements
$v\to v^\ast$, $\sigma\to\tau$, $\tau\to\xi$, and $\xi\to-\zeta$
are made to the latter equation.

In order to give some appreciation of the scaling of
the variables in Eq.\ (\normb), we give more explicit forms
for them in the limit
$\omega_{pi}^2\lsls\omega_0^2\lsls\omega_{pe}^2\lsls\Omega_e^2$
and $T_i\lsls T_e$ (these may be the conditions applicable
as the lower hybrid wave propagates through the low density
part of a tokamak plasma).  Equation (\normb) then becomes
$$\xi={x\over2\sqrt3\lambda_{De}},\quad
\tau={\omega_0t\over2},\quad
\zeta={gz-x\over\sqrt3\lambda_{De}},\quad
v=\left({\epsilon_0\over4n_0T_e}\right)^{1/2}E,\eqn(\en\norma)$$
where $\lambda_{De} = v_{te}/\omega_{pe}$ is the Debye length
and $g=\omega_0/\omega_{pe}$.

\section \basic. Basic Properties of the Equation.

Equation (\eq) is invariant under the scaling transformation
$$\xi\to\lambda^{-3}\xi,\qquad\tau\to\lambda^{-2}\tau,\qquad
\zeta\to\lambda^{-1}\zeta,\qquad v\to\lambda v.\eqn(\en\scaling)$$
This invariance allows us, for example, to normalize
the scale length of the ray in the $\zeta$ direction to
unity.

If we set $\partial/\partial\tau =0$ in Eq.\ (\eq), we obtain,
as expected, the complex modified Korteweg--deVries equation
as the equation obeyed by steady-state fields.  If $\partial
/\partial\xi=0$, we obtain the nonlinear Schr\"odinger
equation, which is soluble by the inverse scattering
method.\ref\scott\  Unfortunately, not much use can be made of this
fact because, as we shall see, the imposition of the correct
boundary conditions involves solving Eq.\ (\eq) in a finite
domain in $\xi$.

For each of the four conservation laws obtained for the
complex modified Korteweg--deVries equation, there is
a corresponding conservation law for Eq.\ (\eq).  These
are given in Table \constab.
The second and third of these laws are
statements of the conservation of momentum and
energy respectively.  We will return to these shortly.

In order to determine the correct boundary
conditions for Eq.\ (\eq), we Fourier transform in $\zeta$.
Defining the Fourier transform of $v$ by
$$V(\xi,\kappa,\tau)=\int_{-\infty}^\infty
v(\xi,\zeta,\tau) \exp(-i\kappa\zeta) \,d\zeta$$
we obtain
$$\eqalignno{
V_\tau + c V_\xi + i\Omega V + i N(V)&= 0,&(\en\eqtx a)\cr
V(\kappa=0)&=0,&(\+b)\cr}$$
where $c= 1/\kappa$, $\Omega = \kappa^2$, and
$N(V)$ is the Fourier transform of $-\abs v^2 v$.  This
equation is hyperbolic in $(\tau,\xi)$ space.  However,
the direction of the characteristic velocity
$c$ depends on $\kappa$.  This means that we must specify
boundary conditions at each end of a strip in $\xi$.
More precisely, the full initial
and boundary conditions required for solving Eq.\ (\eqtx) in the
domain
$$\xi_0<\xi<\xi_1,
\qquad -\infty<\zeta<\infty,
\qquad \tau_0<\tau<\infty$$
are
$$V(\xi_0<\xi<\xi_1,\kappa,\tau_0),
\quad V(\xi_0,\kappa>0,\tau>\tau_0),
\quad V(\xi_1,\kappa<0,\tau>\tau_0).\eqn(\en\bc)$$

Taking the second and fourth conservation
laws in Table \constab, integrating them over $\zeta$
and $\xi$, and transforming them to $\kappa$ space,
we obtain the laws of conservation of momentum
and energy as they apply to the domain of the
problem
$$\eqalignno{
{d\over d\tau}\int_{\textstyle \xi_0}^{\textstyle \xi_1}
\int_{-\infty}^\infty \Mscr\,d\kappa\,d\xi
+\left.\int_{-\infty}^\infty \Fscr\,d\kappa
\right|_{\textstyle \xi=\xi_0}^{\textstyle \xi_1}&=0,&(\en\consm)\cr
{d\over d\tau}\int_{\textstyle \xi_0}^{\textstyle \xi_1}
\int_{-\infty}^\infty \Escr\,d\kappa\,d\xi
+\left.\int_{-\infty}^\infty \Pscr\,d\kappa
\right|_{\textstyle \xi=\xi_0}^{\textstyle \xi_1}&=0,&(\en\conse)\cr}
$$
where
$\Mscr=\kappa\abs V^2$ is the spectral momentum density,
$\Fscr=c\Mscr$ is the spectral force density,
$\Escr=\abs V^2$ is the spectral energy density,
and $\Pscr=c\Escr$ is the spectral power density.
The characteristic velocity $c$ is
the group velocity of the lower hybrid waves.  The
boundary conditions given in Eq.\ (\bc)
stipulate that we should specify the waves entering the
domain of the problem, but not those that leave this domain.

Given that we are treating a nonlinear problem, it may be a
little surprising the boundary conditions are applied in the
same way as for the corresponding linear problem.  The mathematical
reason for this is that the nonlinear term does not involve any
$\tau$ or $\xi$ derivatives.  This arises because,
in deriving Eq.\ (\eq), we assumed that the
nonlinear term was very small.  The expressions for the energy
density, power density, etc.\ are therefore correctly given by
linear theory.  Although the nonlinearity is small, it can have
a finite effect because the ordering also assumes that the
length over which the interaction takes place is large.

Notice that the nature of the equation has forced us to
take boundary conditions which are in accord with physical
notions of power flow.  This has happened because we have
the time variation in Eq.\ (\eq).  If we consider the steady-state
equation, the complex modified Korteweg--deVries equation,
which is obtained by setting $\partial/\partial\tau=0$
in Eq.\ (\eq)
$$v_\xi-v_{\zeta\zeta\zeta} - (\abs v ^2 v)_\zeta = 0,\eqn(\en\eqss)$$
then we would clearly like to impose the same boundary conditions
as for Eq.\ (\eq) namely Eq.\ (\bc).
But the concept of power flow cannot be defined from
Eq.\ (\eqss) and so there is no guarantee that the
boundary conditions given in Eq.\ (\bc) are sensible.  Indeed,
the evidence of Ref.\ \karney\ suggests
that Eq.\ (\eqss) is ill-posed
with these boundary conditions.  This can be so because,
even if Eq.\ (\eq) is given steady-state boundary conditions, i.e.,
$$V(\xi_0<\xi<\xi_1,\kappa,\tau_0\to -\infty),
\quad {\partial\over\partial\tau} V(\xi_0,\kappa>0,\tau) = 
      {\partial\over\partial\tau} V(\xi_1,\kappa<0,\tau) = 0,$$
it is not necessarily the case that the solution to Eq.\ (\eq)
approaches a steady state.

As mentioned in Sec.\ \deriv, the ordering we used in
deriving Eq.\ (\eq) was a maximal ordering.  We can eliminate
some of the terms in Eq.\ (\eq) using a subsidiary ordering.
E.g., the linear limit is obtained by letting
$v\to\delta\,v$ and taking the limit $\delta\to 0$.  Similarly,
the non-dispersive limit is obtained by letting
$\zeta\to\zeta/\delta$, $\xi\to\xi/\delta$ and again taking
the limit $\delta\to 0$.  It is instructive to examine
this case in more detail because there are indications
that if $\xi_1-\xi_0$ is sufficiently large then the dispersive
term must be included.  Consider the steady-state
equation (\eqss) without the
dispersive term $v_{\zeta\zeta\zeta}$.
Multiplying this equation by $v^\ast$ and adding the complex conjugate
(these are the same operations that lead to the
second conservation law in Table \constab), we obtain
$$u_\xi + uu_\zeta = 0\eqn(\en\shock)$$
where $u = 3\abs v^2$.  Since the boundary conditions
involve $\arg v$, Eq.\ (\shock) should be supplemented by
another equation.  However, this is not so
if we impose initial conditions $v(\xi=\xi_0,\zeta)$.
As is well known, the solution develops a shock
at $\xi=\xi_0+1/\max[du(\xi=\xi_0,\zeta)/d\zeta]$, where
$\partial u/\partial\zeta$ is infinite.  Before this point
is reached the ordering assumed for the independent variables,
i.e., that $\partial/\partial\zeta=O(\delta)$, breaks down.
Inclusion of the dispersive term prevents the shock from forming.
Indeed, Eq.\ (\eqss) when treated as an initial value problem
in $\xi$ is well-posed in an infinite domain in $\xi$.
Although Eq.\ (\eq) with the correct boundary conditions is
different from Eq.\ (\eqss) with initial conditions, it is likely
that Eq.\ (\eq) is also ill-posed if the dispersive term is
ignored and if the width of the domain in $\xi$ is large enough.

\section \numerics.  Numerical Results.

We investigate the solution to Eq.\ (\eq) by numerically
solving Eq.\ (\eqtx) with boundary and initial conditions
$$\eqalignno{V(0<\xi<\xi_1,\tau=0)&=0,&(\en\bca a)\cr
V(\xi=0,\kappa>0,\tau>0)&
=v_0\pi\,u(\kappa-\kappa_0)(\kappa-\kappa_0)
\exp[-\fract1/4 (\kappa-\kappa_0)^2],&(\+ b)\cr
V(\xi=\xi_1,\kappa<0,\tau>0)&=0,&(\+ c)\cr}$$
where $u$ is the unit step function,
$v_0$ and $\kappa_0$ are real constants and $\kappa_0\geq 0$.
The method of solution is given in Appendix \appnum.
Without loss of generality, we have taken $\tau_0=\xi_0=0$.
In $\zeta$ space Eq.\ (\bca b) becomes
$$\left.v(\xi=0)\right|_{\kappa>0}
=v_0[1+\zeta Z(\zeta)]\exp(i\kappa_0\zeta)$$
where $Z$ is the plasma dispersion function.  With this form of the
boundary conditions, we now have three parameters to vary:  $v_0$,
$\kappa_0$, and $\Delta\xi\eqv\xi_1-\xi_0=\xi_1$.
The geometry of the waveguide array which might
give this input spectrum will be discussed in Sec.\ \thresh.

We begin by considering the linear non-dispersive limit;
i.e., we remove the third and fourth terms in Eq.\ (\eq a).
Figure \linfig\ shows the solution at various
times.  Early in the evolution of the waves, only the
lowest $\kappa$ components have propagated throughout the
plasma because these have the greatest group velocities.
Later on, the other $\kappa$ components have had time to
reach the far end of the system, and the steady state
is attained.  As $\tau\to\infty$, we have $v = v(\zeta)$.
In the original $(x,z)$ coordinate system,
this corresponds to the familiar propagation along resonance
cones.

In order to further diagnose the results, we identify
three sets of waves,
the incident waves ($\xi=0$ and $\kappa>0$),
the reflected waves ($\xi=0$ and $\kappa<0$),
and the transmitted waves ($\xi=\xi_1$ and $\kappa>0$).
With our choice of boundary conditions the
incident waves are constant for $\tau > 0$ and there
are no waves incident at the boundary $\xi=\xi_1$.
We define reflection and
transmission coefficients, $R$ and $T$,
as the ratios of the instantaneous
power reflected and the instantaneous power transmitted to
the incident power.  These powers are defined
by integrating $\Pscr$ over the appropriate segment of
the boundary.
Note that, because energy can be
stored in the plasma, we need not have $R+T=1$, although
this is true in a time-averaged sense.  Furthermore, it
is possible that $R$ or $T$ can momentarily exceed unity.
In order to determine where in $\kappa$ space the power
is concentrated, we define $\ave\kappa$ for each
wave component as the ratio
of the force to the power for that component.  (Since
$\Pscr=\abs V^2/\kappa$ and $\Fscr=\abs V^2$,
this ratio has the dimensions of $\kappa$.)

For the case shown in Fig.\ \linfig, $R$ is obviously zero
and the average $\kappa$ of the reflected waves, $\ave\kappa_r$,
is undefined.  Figure \linrefl\ shows $T$ and the average
$\kappa$ of the transmitted waves, $\ave\kappa_t$, for this case.
The transmission coefficient $T$ rises in steps as
each $\kappa$ mode reaches the far boundary.
[Because of the use of periodic boundary conditions
in the numerical solution of Eq.\ (\eq), there are
only a discrete set of $\kappa$ modes.]
Similarly, $\ave\kappa_t$ rises with time since modes
with higher values of $\kappa$ take longer to
traverse the domain.  As $\tau\to\infty$, $T$ approaches
unity, signifying total transmission, and $\ave\kappa_t$
approaches $(\pi/2)^{1/2}\approx 1.25$ which is the same as the
average value of $\kappa$ for the incident waves, $\ave\kappa_i$.

We now turn to the nonlinear dispersive problem.  Because we
expect that the effect of the nonlinearity will
increase with both $v_0$ and $\Delta\xi$, we begin with a case
where these quantities are chosen small enough that the
nonlinearity only weakly modifies the propagation.  Figure
\aiii\ shows the solution for
such a case with $v_0=3$, $\kappa_0=0$,
and $\Delta\xi=0.1$.  It is found that $v$ attains a steady state
by $\tau=2$ approximately and only this state is shown
in Fig.\ \aiii.  The evolution to this steady state is
similar to that for the linear equation.
In Fig.\ \aiiirefl\ are displayed the
reflection and transmission coefficients.  Note that the
reflection coefficient settles down to a small value (somewhat
less than 2\%).  

As the amplitude of the boundary value $v_0$ is increased,
the solution exhibits quite different behavior from the
linear solution.  In Fig.\ \aiv, we have chosen the same
parameters as in Fig.\ \aiii, except that $v_0$ has been increased
to 4.  Now $v$ appears never to reach a steady state, but
instead oscillates in some aperiodic fashion.  This
can be seen in Fig.\ \aivrefl, where $R$, $T$, $\ave\kappa_r$,
and $\ave\kappa_t$ are plotted.  From Fig.\ \aiv, we can
see what is happening during these oscillations.
Various $\kappa$ components in the forward wave
nonlinearly interact to produce a reflected wave [Fig.\ \aiv(b)].
This interaction causes a severe depletion of the low $\kappa$
components of the forward wave and transfer of this energy
into higher $\kappa$ components of the forward wave and
into the reflected wave [Fig.\ \aiv(c)].  After the interaction
has nearly gone to completion, the nonlinearly excited components
of the field transit out of the system and the fields relax to
a state in which there is little variation in $\abs V$ with $\xi$
[Fig.\ \aiv(d)].  The nonlinear interaction then begins again
and the cycle approximately repeats itself [Fig.\ \aiv(e)].
In $\zeta$ space,
this interaction is manifested by a narrowing and peaking of the
electric field amplitude $v$.  A typical field pattern is shown in
Fig.\ \aiv(f).

Returning to Fig.\ \aivrefl, we see
that the reflection is appreciable, oscillating around about
20\%.  Correspondingly, $T$ oscillates around 80\%.  
Because the energy stored in the field comes out in bursts,
$T$ is occasionally greater than unity.  Finally, we note
that $\ave\kappa_t$ exceeds $\ave\kappa_i\approx 1.25$
by approximately a factor of 2.

Because the long-time solution of the fields does not reach a
steady state, it is useful to describe the solution in terms of
its temporal spectrum.  To do this, we take a time record of the
transmitted and reflected waves for $\tau_a<\tau<\tau_b$.  We
choose $\tau_a$ to be large enough for the transients to have
disappeared from the solution and $\tau_b-\tau_a$ to be large
enough to include several oscillations of the solution.
Effects arising from the finite record length $\tau_b-\tau_a$
are partially eliminated by multiplying by a cosine
window $\half\{1-\cos[2\pi(\tau-\tau_a)/(\tau_b-\tau_a)]\}$.
The resulting function is then transformed in time using
a discrete Fourier transform to give the output spectrum
$V(\xi^\prime,\kappa,\sigma)$
where $\sigma$ is the slow frequency variable (conjugate
with $\tau$),
$\xi^\prime=\xi_1$ for $\kappa>0$ (the transmitted wave),
and $\xi^\prime=0$ for $\kappa<0$ (the reflected wave).
[We assume a space-time dependence of
$\exp(i\kappa\zeta-i\sigma\tau)$.]

In Fig.\ \aivspec, we plot the output spectrum $V(\xi^\prime,\kappa,\sigma)$
for the case shown in Figs.\ \aiv\ and \aivrefl.  The spectrum
is computed between $\tau_a=1$ and $\tau_b=5$.  The large peaks at
$\sigma=0$ are the steady-state components of the transmitted and
reflected waves.  The broad spectrum in $\sigma$ is symptomatic of
the aperiodic nature of the waves.
There is an interesting correlation
in the spectrum:  the positive frequency ($\sigma$) components of
both the transmitted and reflected waves tend to have
larger values of $\abs\kappa$ than the negative frequency
components.

We have seen that the solution becomes more turbulent and
that the reflection increases as $v_0$ is increased.  The
other parameters that may be varied are $\Delta\xi$ and $\kappa_0$.
We now investigate the influence of these parameters.

We begin by varying $\Delta\xi$.
Figure \xv\ shows the output spectrum for $v_0=4$, $\kappa_0=0$
and $\Delta\xi=0.05$.  We see that there is almost no energy in
the components of the wave with $\sigma\neq0$;  i.e., the fields
nearly reach a steady state.  The reflection coefficient is
smaller than in the previous case (where $\Delta\xi=0.1$) being
approximately 4\%.  Similarly $\ave\kappa_t$ is only slightly
greater than $\ave\kappa_i$ (1.4 as compared to 1.25).
On the other hand, if we increase the value of $\Delta\xi$
to 0.2 (twice its value in Fig.\ \aivspec), a broad turbulent
spectrum is recovered (Fig.\ \xii).  Although $R$ and $\ave\kappa_t$
have approximately the same values as for $\Delta\xi=0.1$,
there is now more energy in the non-steady components of the
output spectrum.

Figures \ki\ and \kii\ show the effect on the output spectrum
of increasing $\kappa_0$ (keeping $v_0=4$ and $\Delta\xi=0.1$).
In Fig.\ \ki, where $\kappa_0=1$,
we see that spectrum consists of a few narrow bands indicating
that the solution is nearly periodic.  Furthermore, the reflected
wave consists almost entirely of negative frequency components.
The reflection coefficient for this case is about 12\%, or half of
what it was with $\kappa_0=0$.  The increase of $\ave\kappa_t$ over
$\ave\kappa_i$ is moderate (about 3 as compared to 2.42).
Increasing $\kappa_0$ to 2 (Fig.\ \kii), we see much the same
picture except that the frequency of the oscillations is higher.
The reflection coefficient of is further reduced to 7\%\ and
there is now little difference between $\ave\kappa_t$ and
$\ave\kappa_i$.

\section \thresh.  Threshold for the Nonlinearity.

In the case shown in Figs.\ \aiv--\aivspec, the nonlinearity causes
three phenomena of importance to lower hybrid heating experiments:
(1) the reflection coefficient is appreciable;
(2) the solution reaches a turbulent state;
(3) the mean wavenumber of the waves transmitted into
the plasma is increased.
Comparing the results presented here with those of
Sec.\ VIII of Ref.\ \karney\ in which the solution of the steady-state
problem was attempted, we see that the conditions for the occurrence
of these phenomena closely agree with the conditions under which the
steady-state problem had no solution and reflection was large.

Before quoting the threshold results from Ref.\ \karney, we
use the scaling invariance Eq.\ (\scaling) to
rewrite the boundary condition Eq.\ (\bca b) as
$$V(\xi=0,\kappa>0,\tau>0)
=v_0\pi\,u(\kappa-\kappa_0)(\kappa-\kappa_0)\,\Delta\zeta^2
\exp[-\fract1/4 \Delta\zeta^2(\kappa-\kappa_0)^2].$$
In real space, we have
$$\left.v(\xi=0)\right|_{\kappa>0}
=v_0[1+(\zeta/\Delta\zeta) Z(\zeta/\Delta\zeta)]
\exp(i\kappa_0\zeta).$$
This roughly corresponds to excitation of lower hybrid
waves by a waveguide array of width $\Delta\zeta$.  The amplitude
of the electric field in the plasma is $v_0$ and the phasing
of the waveguides is such as to give a wavenumber of $\kappa_0$.
Thus, if the waveguides are phased $0,\pi,0,\pi,\ldotss$,
we have $\kappa_0=\pi M/\Delta\zeta$ where $M$ is the number of
waveguides.  [The average wavenumber of the spectrum
of a single lower hybrid ray is
$\ave\kappa_i$ which, for $M$ large,
is approximately $\kappa_0+2(2/\pi)^{1/2}/\Delta\zeta$.  The
first term is attributable to the phasing of the waveguides and
the second term arises because of the finite width of the
waveguide array.]

For these boundary conditions, the threshold conditions given in
Ref.\ \karney\ are
$$\eqalignno{v_0&\grapprox2\kappa_0,&(\en\cond a)\cr
\Delta\xi&\grapprox\sqrt2v_0^{-3}.&(\+b)\cr}$$
The latter conditions is only correct for $M$ small ($\lsapprox4$).
For larger values of $M$, it should be replaced by
$$\Delta\xi\grapprox\Delta\zeta\, v_0^{-2},\eqn(\+c)$$
which specifies that $\Delta\xi$ should exceed the distance
for a shock to form in the dispersionless
equation (\shock).  Equation (\cond) gives the conditions
under which the three phenomena described at
the beginning of this section occur.

If we now undo the normalizations using Eq.\ (\norma),
we find that Eq.\ (\cond) becomes
$$\eqalignno{E_{z0}\,\delta z/T_e&\grapprox 4\sqrt3\pi,
&(\en\conda a)\cr
\Delta x &\grapprox 4\sqrt6 (T_e/E_{x0})/\beta,&(\+b)\cr
\Delta x &\grapprox g\,\Delta z/\beta,&(\+c)\cr}$$
where $E_{x0}$ and $E_{z0}$ are the electric field amplitudes
perpendicular and parallel to the magnetic field (so that
$E_{z0}=gE_{x0}$), $\Delta z$ is the width of the
waveguide array, $\delta z=\Delta z/M$ is the width of a single
waveguide (assuming a $0,\pi,0,\pi,\ldotss$ phasing), $\Delta x$
is the width of the nonlinear region of the plasma in the $x$
direction, and $\beta$ is $\fract1/4\epsilon_0E_{x0}^2/n_0T_e$,
the ratio of the electric field energy to the plasma kinetic
energy.

\section \conclusions.  Discussion.

We have examined the nonlinear evolution of a lower hybrid
wave in two dimensions and time.  Under the conditions given
in Eq.\ (\conda), the nonlinearity can cause appreciable
reflection, turbulent variation in the fields, and an
increase in the wavenumbers of the transmitted waves.  In
Ref.\ \karney, it was found that these conditions could be
satisfied in a small laboratory plasma or near the edge of
a tokamak plasma.  In such circumstances, the
assumption of a steady state and the analyses in
Refs.\ \moralesa--\karney\ based on this assumption are wrong.

The most important application of lower
hybrid waves is for heating a tokamak plasma to ignition
temperatures.  It is therefore important to understand the
processes that may modify the lower hybrid waves before
they have penetrated to the center
of the plasma.  Unfortunately, it is not possible to obtain
quantitative estimates of these effects with the theory as
outlined in the preceding sections.  The reason is that, as
we have pointed out, the nonlinearity is only important in
a tokamak near the edge of the plasma and in that region
many other physical processes are likely to be involved.
Examples of effects which should probably
be included to give a full understanding of lower hybrid wave
propagation near the edge are: nonlinear coupling to the
second (left-going) ray; electromagnetic effects; density
and temperature gradients; saturation of the nonlinearity;
ion inertia in the low frequency equations; coupling to
low-frequency drift waves.

Although Eq.\ (\eq) should be regarded as a very simplified model
equation describing the propagation of lower hybrid waves near
the edge of a tokamak plasma, some of the phenomena predicted
by this equation have been observed in
the lower hybrid heating experiment on
Alcator--A.\ref{\schuss, \surko}
The CO$_2$-laser scattering diagnostic on
that experiment indicated a
broadening of the frequency spectrum of the waves.  This is
consistent with the turbulent spectra seen in the solutions
of Eq.\ (\eq).  Furthermore, the spectrum becomes asymmetric as
the wave propagates into the plasma, the peak of the spectrum
being shifted down from the frequency of the injected waves.
This may be a result of the observation in Sec.\ \numerics\ that
the components of the transmitted and reflected
waves which are down-shifted in frequency have a smaller
(in absolute value) wavenumber than those which are up-shifted.
If those waves with higher wavenumbers are preferentially
damped as the wave travels into the plasma (e.g., by
electron Landau damping), we would expect to see
a net downwards shift in the spectrum of waves.

The other somewhat puzzling result of the Alcator--A
experiment was the apparent independence of the heating
results to the phasing of the waveguides.  In addition,
in order to explain the density dependence of the heating
results, it was postulated that the $n_z$ (parallel
index of refraction $k_zc/\omega$) spectrum of the waves
was shifted to being peaked at around $n_z=5$ irrespective
of the phasing of the waveguides.  (This is to be compared
with values of $n_z$ predicted on the basis of
linear theory of less than 1.5 for the waveguides in phase
and of 3 for the waveguides out of phase.)  Again, this is
qualitatively in agreement with the theoretical results
of Sec.\ \numerics.  There, the average wavenumber of
the waves transmitted from the nonlinear region into
the center of the plasma was roughly twice
that of the wave injected into the nonlinear region
when $\kappa_0=0$ (corresponding to the wave guides being in phase).
As the phase difference between neighboring waveguide
becomes finite (i.e., as $\kappa_0$ is increased), the
amount by which the wavenumber spectrum is shifted is reduced.

It is not clear whether the nonlinear reflection predicted by the
theory given in this paper would be observed as an increase in the
reflected power measured in the waveguides.  It may be that this
power is reflected again on the cutoff at $\omega=\omega_{pe}$ very
close to the plasma edge.  This would convert the power into the
other lower hybrid ray.

Thus, it appears that the physics included in Eq.\ (\eq) may be
responsible for some of the results of the Alcator--A experiment.
In order to be able to say definitively whether or not the experimental
observations are a result of the processes included in Eq.\ (\eq),
two advances are required.  Firstly, the theory has to be refined
to a point where quantitative comparisons are possible.  Secondly,
additional experimental observations are required.  (At present,
Alcator--A is the only tokamak experiment for which the
detailed observations provided by the CO$_2$-laser scattering
diagnostic are available.  In addition, it would be helpful
to have accurate measurements of
the plasma conditions near the edge of the plasma.)

We should point out that other theories have
beeen advanced to explain the Alcator--A results.  In particular,
Bonoli and Ott\ref\bonoli\ have suggested a linear theory.  This is
supported by the observed linear dependence between the density
fluc\-tu\-a\-tions and the applied power (over a fairly large range)
which suggests either that the phenomena are linear or else that
the nonlinear processes saturate at a low amplitude.

Another interesting theoretical result, given by
Morales,\ref\moralesb\ concerns the
coupling of rf energy into lower hybrid waves at the plasma
edge.  A density gradient was included in his model and
a temporal evolution was allowed.  As in the theory presented
here, it was found that a steady state need not be reached;
rather, the rf energy entered the plasma in bursts
similar to the behavior seen in Fig.\ \aiv.  However,
only a single $n_z$ component was included so that the
nonlinear coupling of different $n_z$ components was
disallowed.

In summary, we have presented a theory of the nonlinear
propagation of lower hybrid waves.  At sufficiently large
powers, the fields become turbulent and the wavenumber
spectrum is shifted upwards.  The theory should
accurately describe the propagation in small laboratory
devices.  While we may expect qualitatively similar results
near the edge of a tokamak plasma, other physical effects
need to be included to obtain an accurate description
of the lower hybrid fields in this region.

\acknowledgments

I wish to thank my collaborators on Ref.\ \karney, F. Y. F. Chu and
A. Sen, who encouraged this extension of our earlier work and who
helped in its formulation.  I would also like to thank N. J. Fisch
for extensive discussions and
R. H. Berman for his comments on a preliminary version of
this paper.

This work was supported by the U.S. Department
of Energy under Contract DE--AC02--76--CH03073.

\appendix \appnum.  Numerical Procedure for Solving {Eq.\ (\eq)}.

The numerical solution of Eq.\ (\eq) is
carried out in Fourier space so that Eq.\ (\eq) becomes
Eq.\ (\eqtx).  Periodic boundary conditions are used in the
$\zeta$ direction, $v(\xi,\zeta+L,\tau)=v(\xi,\zeta,\tau)$;
therefore, the Fourier spectrum is discrete the spacing between
modes being $\delta\kappa=2\pi/L$.  We use $n$ points to describe
$v$ over a single period $L$.
Transformations between the $\kappa$
and $\zeta$ spaces are achieved using a discrete Fourier
transform.  Thus, $V$ is approximated by $n$ Fourier modes.
The $\xi$ coordinate is approximated by a grid
whose spacing is $\delta\xi$.
This equation is solved by splitting each
time step $\delta\tau$ into two equal pieces and by
approximating Eq.\ (\eqtx a) in the interval $0 \leq
\tau < \delta\tau$ by
$$V_\tau =-2\times\case{
c V_\xi,&\for 0\leq\tau<\half\delta\tau,\cr
i\Omega V + i N(V),&\for \half\delta\tau\leq\tau<\delta\tau.\cr}
\eqn(\e)$$
(For simplicity, we describe the solution only for the first
time step.  The extension beyond this is obvious.)

In each half time step this equation is a partial differential
equation in only two independent variables.  During the first
half time step, $\kappa$ is merely a parameter and we have
a simple linear wave equation to solve.
We approximate $V^\prime\eqv
V(\tau=\half\delta\tau)$ as given by
Eq.\ (\+) by shifting $V(\tau=0)$
over $C=c\,\delta\tau/\delta\xi$ grid positions.  This
step is exact if $C$ is an integer.  Normally, this is not the
case and in that event we interpolate between neighboring grid
points.  We use linear interpolation on the quantities $\abs V^2$
and $\abs V^2 \arg V$.  This ensures that in this step energy
and momentum are conserved.  The reason for interpolating in
$\abs V^2 \arg V$ rather than $\arg V$ is that in the former case
the ambiguity of the argument of $V=0$ is irrelevant.  The
grid positions within $C$ of the boundary are set to the boundary
value.

During the second half time step we have to solve the nonlinear
Schr\"odinger equation at each position $\xi$.
The method is similar to that used in Ref.\ \karney\ to solve
the steady-state equation (\eqss).  The dispersive term is
treated exactly and the nonlinear term is treated with
a second-order Runge--Kutta scheme.  Thus, $V(\tau=\delta\tau)$
is approximated by
$$\eqalign{
V^{\prime\prime}&=BV^\prime+DN(V^\prime),\cr
V(\delta\tau)
&=BV^\prime+\half D[N(V^{\prime\prime})+N(V^\prime)],\cr}$$
where $B=\exp(-i\Omega\,\delta\tau)$ and $D=-(1-B)/\Omega$.  The
nonlinear term $N(V)$ is calculated
by transforming $V$ into $v$ using the
discrete Fourier transform, computing $-\abs v^2v$, and transforming
back into $\kappa$ space.  In order to avoid problems of aliasing
in the computation of $N(V)$, the highest $n/2$ Fourier modes are
artificially set to zero.

The accuracy of the numerical integration is checked using
the momentum- and energy-conserv\-ation laws.
Specifically, the time integrals of  Eqs.\ (\consm)
and (\conse) are numerically computed.  These
are divided by the total momentum input into the
plasma (at $\xi=\xi_0$) and by the total energy
injected into the plasma (both at $\xi=\xi_0$, $\kappa
> 0$ and at $\xi=\xi_1$, $\kappa < 0$) to provide two
measures of the accuracy of
the numerical integration, $\Delta\Mscr$ and $\Delta\Escr$.

The results presented in Sec.\ \numerics\ were (with the
exception of Figs.\ \linfig\ and \linrefl) computed with
$n=2^8$, $L=20$, and $\delta\xi=10^{-3}$.  The time step
was taken to be $\delta\tau=10^{-3}$ in Figs.\ \aiv--\aivspec,
and \xii, $\delta\tau=2\times10^{-3}$
in Figs.\ \xv, \ki, and \kii, and $\delta\tau=5\times10^{-3}$
in Figs.\ \aiii\ and \aiiirefl. The error parameters
$\Delta\Mscr$ and $\Delta\Escr$ were less than about
$5\times10^{-4}$ at $\tau=5$ in all these cases.

The case shown in Figs.\ \aiv--\aivspec\ was also computed
with $n=2^9$, $L=20$, and $\delta\tau=\delta\xi=5\times10^{-4}$
(i.e., the grid spacing was halved in $\tau$, $\zeta$, and $\xi$).
The solution agreed well with the solution obtained using the
coarser grid until about $\tau=1.5$.  Thereafter, the solutions
diverged from each other.  This is to be expected in a system
which exhibits turbulent solutions because the solution is
typically very sensitive to the initial conditions.  Numerical
errors, which have an effect similar to changing the initial
conditions slightly, can therefore lead to large changes in the
solution.  However, although the detailed solution is different
after $\tau=1.5$ in these two cases,
the general character of the solution is the same.  Thus,
the output spectrum for the case with the finer grid spacing
shows the same features as the spectrum given in Fig.\ \aivspec.

\endpaper
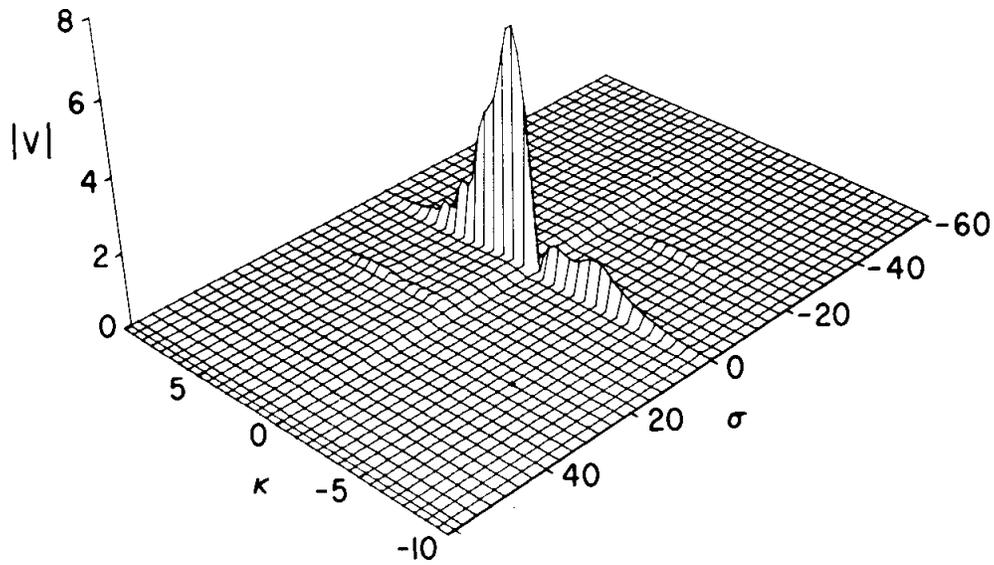
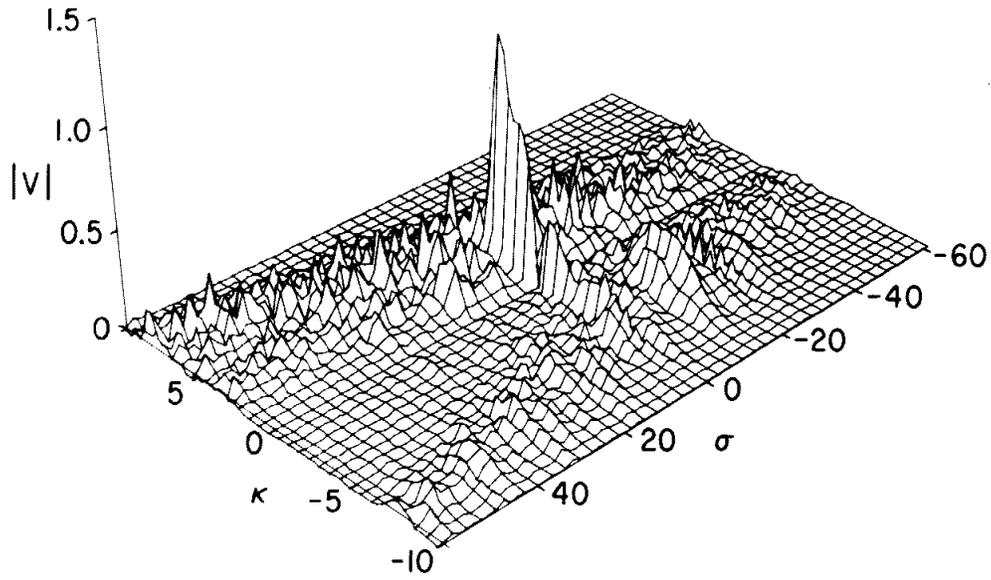
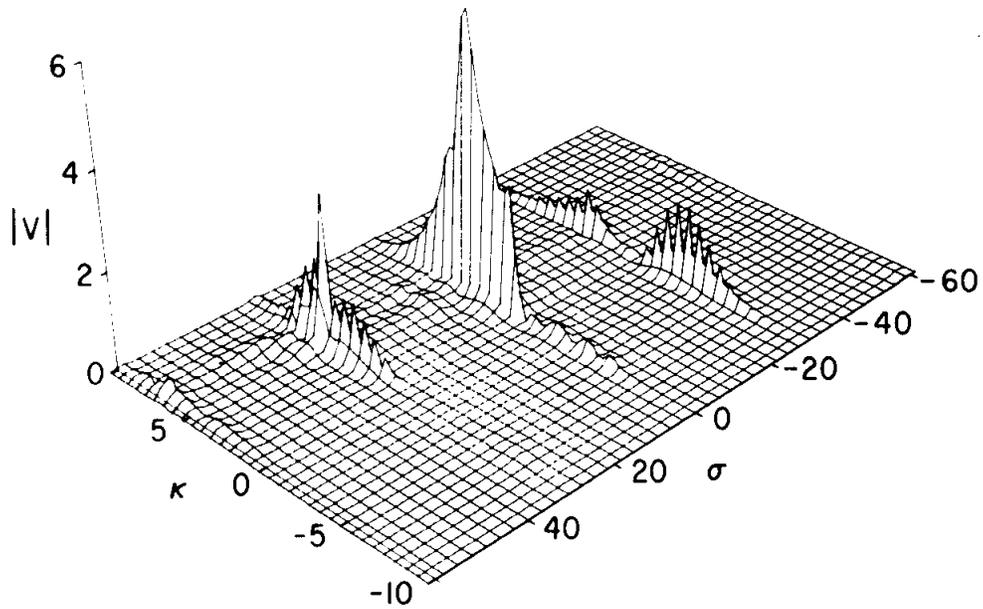
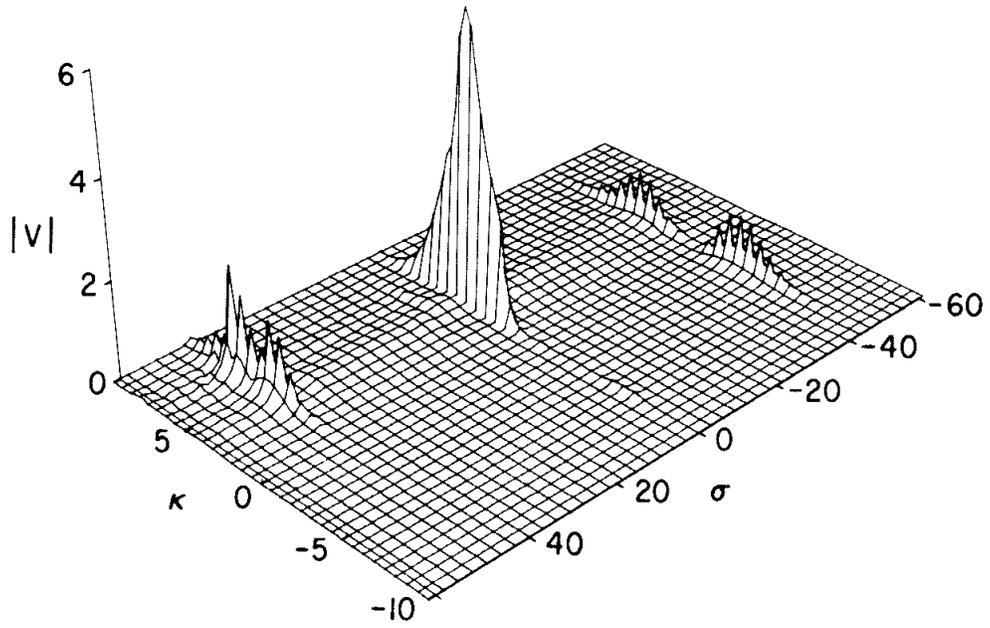
